\documentclass[nofootinbib,showpacs,a4paper,twocolumn,10pt]{revtex4-1}
\usepackage[utf8x]{inputenc}
\usepackage{amsmath}
\usepackage{amsfonts}
\usepackage{amssymb}
\usepackage{graphicx}
\usepackage{subfigure}

\usepackage{amsmath}
\usepackage{amsfonts}
\usepackage{amssymb}
\usepackage{amsthm}
\usepackage{mathrsfs}
\usepackage{dsfont}

\DeclareMathAlphabet{\mathpzc}{OT1}{pzc}{m}{it}

\newcommand{\Laplacian}{\ensuremath{\triangle}}

	\newcommand{\Prop}{\sim}

	\newcommand{\Ket}[1]{\left|#1\right\rangle}
	\newcommand{\Bra}[1]{\left\langle#1\right|}

	\newcommand{\BK}[2]{\left\langle#1|#2\right\rangle}
	\newcommand{\BAK}[3]{\Bra{#1}#2\Ket{#3}} 
	 
	\newcommand{\KB}[2]{\left|#1\right\rangle\!\left\langle #2\right|}

	\newcommand{\V}[1]{\ensuremath{\boldsymbol{#1}}}			

	\newcommand{\mr}[1]{\mathrm{#1}}			

	\newcommand{\br}[1]{\left( #1 \right)}
	\newcommand{\brr}[1]{\left[ #1 \right]}
	
	\newcommand{\of}[1]{\!\br{#1}}

	\newcommand{\sbr}[1]{( #1 )}
	\newcommand{\sbrr}[1]{[ #1 ]}
	
	\newcommand{\sof}[1]{\!\sbr{#1}}

	\newcommand{\Sum}[2]{\sum\limits_{#1}^{#2}}

	\newcommand{\Int}[3]{\int\limits_{#1}^{#2}\mr{d}#3\,}
	\newcommand{\sSum}[2]{\sum_{#1}^{#2}}

	\newcommand{\sInt}[3]{\int_{#1}^{#2}\mr{d}#3\,}

	\newcommand{\EA}[1]{\xpc{#1}}

	\newcommand{\xpc}[1]{\left\langle #1 \right\rangle}

	\newcommand{\sEA}[1]{\sxpc{#1}}

	\newcommand{\sxpc}[1]{\langle #1 \rangle}

	\newcommand{\Sphere}{\ensuremath{\mathbb{S}}}
	\newcommand{\Reals}{\ensuremath{\mathbb{R}} }

	\newcommand{\Integers}{\ensuremath{\mathbb{Z}} }

	\renewcommand{\d}{\mathrm{d}}

	\newcommand{\Landau}[1]{\mathpzc{O}\of{#1}}
	\newcommand{\landau}[1]{\mathpzc{o}\of{#1}}

		\newcommand{\Min}[2]{\min\of{#1,#2}}
		\newcommand{\Max}[2]{\max\of{#1,#2}}
		\newcommand{\Abs}[1]{\left\vert #1 \right\vert}
		\newcommand{\sAbs}[1]{\vert #1 \vert}

		\newcommand{\Id}{\mathds{1}}
		
		\newcommand{\HGF}[4]{{ }_{2}F_{1}\left[ #1 \bigg| \begin{array}{c} \tiny{(#2)} , \tiny{(#3)} \\ \tiny{(#4)}  \end{array} \right]}

	\newcommand{\Meter}{\mr{m}}
	\newcommand{\Sec}{\mr{sec}}

\begin{document}

	\title{Effective medium approximation for lattice random walks with long-range jumps}

	\author{Felix Thiel}
	\email{thiel@posteo.de}
	\affiliation{Institut f\"ur Physik, Humboldt-Universit\"at zu Berlin, Newtonstr. 15, 12489 Berlin, Germany}
	\author{Igor M. Sokolov}
	\affiliation{Institut f\"ur Physik, Humboldt-Universit\"at zu Berlin, Newtonstr. 15, 12489 Berlin, Germany}


	\begin{abstract}
		We consider the random walk on a lattice with random transition rates and arbitrarily long-range jumps.
		We employ Bruggeman's effective medium approximation (EMA) to find the disorder averaged (coarse-grained) dynamics.
		The EMA procedure replaces the disordered system with a cleverly guessed reference system in a self-consistent manner.
		We give necessary conditions on the reference system and discuss possible physical mechanisms of anomalous diffusion.
		In case of a power-law scaling between transition rates and distance, lattice variants of L\'evy-flights emerge as the effective medium, and the problem is solved analytically, bearing the effective anomalous diffusivity.
		Finally, we discuss several example distributions, and demonstrate very good agreement with numerical simulations.
	\end{abstract}
	\maketitle

	\section{Introduction}
		Anomalous diffusion is a random transport phenomenon characterized by a non-linear growth of the typical dispersion length.
		The dispersion length can be identified with the mean squared displacement, when it is finite.
		Then one has:
		\begin{equation*}
				\xpc{ X^2\of{t} } 
			\Prop
				t^\gamma
			.
		\end{equation*}
		Here $\gamma$ is the characteristic exponent of anomalous diffusion.
		The case $\gamma < 1$ is usually coined subdiffusion, whereas one speaks about superdiffusion, when $\gamma > 1$.
		Superdiffusion appears in plasmas \cite{Liu2008,Zimbardo2015}, diffusive light transmission \cite{Barthelemy2008}, and active particles \cite{Bouzin2015}.
		It appears in the non-physical fields as well: In random searches \cite{Benichou2011,Viswanathan2008}, the motion of living organisms \cite{Othmer1988,Codling2008}, like animals \cite{Edwards2007,Humphries2010}, or humans \cite{Brockmann2006,Gonzalez2008,Song2010,Morris2012}.
		It is also important in epidemic spreading as infected individuals move superdiffusively, \cite{Janssen1999,Hufnagel2004,Brockmann2008,Pastor-Satorras2015}.
		The transport in strongly disordered systems is usually found to be anomalous, see \cite{Bouchaud1990}, however due to different physical reasons that vary from one situation to another, \cite{Thiel2013}.
		In a lattice model, the disorder may be represented by random transition rates, that describe a random walker's jumps between different sites, \cite{Haus1987,Bouchaud1990}.
		Using Arrhenius's law, the rates can be converted into energy differences.
		Hence, such models are known as random-barrier, random-trap, or random-potential models in the physics literature, \cite{Thiel2013,Burov2012,Bustingorry2004,Camboni2012}, but variants are known as random-conductance model in mathematics, \cite{Kumagai2014}, as the master equation also governs the electrical potential of a random resistor network.
		The randomness reflects the possibly vague knowledge about the microscopic dynamics of the diffusing particles.

		Although random master equations pose a rather generic model, they are very difficult to treat analytically.
		It is desirable to consider a disorder-average of the medium that is homogeneous, translationally invariant and can faithfully replace the original disordered system.
		Such an ``homogenization'' procedure is often motivated by the observation that a heterogeneous medium appears homogeneous at larger length scales.
		As the random walker explores more and more of the environment, he will only ``feel'' the average medium.
		Practically, such averages are performed with effective medium approximations (EMA), \cite{Choy1999}. 
		They are numerically very successful and therefore of huge practical relevance, \cite{Nishi2015,Liu2015}.
		One not only uses such concepts to describe transport (especially in the percolation problem), \cite{Kirkpatrick1973,Bustingorry2002,Bustingorry2005}, but also optical phenomena, \cite{Geng2015,Smigaj2008}.

		In principle, the sole necessity for EMA is analytical knowledge about the \textit{effective} topology and propagator.
		Therefore, it has most often been applied to simple-cubic lattices, or other variants with short range transitions, \cite{Haus1987,Kirkpatrick1973}, where a certain jump length threshold can not be exceeded.
		An exception is the work of Parris et al., \cite{Parris1989,Parris2005,Candia2007,Parris2008}.
		Starting late 80's they investigated long-range hopping, although in the context of normal diffusion.
		Later in the 2000's, they considered diffusion on complex networks and focused on traversal times.

		When the nodes of a complex network are embedded in space, i.e. in a spatial network \cite{Barthelemy2011}, long-range connections may arise due to inhomogeneous embedding or due to empirical necessity.
		In a network theoretical treatment, the focus lies on topology; transport is quantified via shortest paths or first passage times.
		Here, we take a different approach: we find an effective topology that exhibits the same behavior in terms of the (anomalous) diffusion constant $K$, whose physical dimensions $\Meter^\mu / \Sec$ also encode the scaling between travel time and displacement.
		This average over the disordered environment restores translational invariance and enables easier analytical treatment of such models.

		EMA can be considered a method of comparison between a disordered model and an \textit{arbitrary} reference model.
		When the correct reference model is chosen, EMA results in a finite effective diffusivity, and in a Markovian description of the disordered system.
		If the effective diffusivity is either zero or infinite, i.e. non-finite, the reference model is not appropriate.
		Either a different reference model needs to be considered or a non-Markovian description has to be employed.
		The main aim of this paper is demonstrating that Bruggeman's variant of EMA, \cite{Bruggeman1935}, can easily be applied to models with more than just nearest-neighbor transitions.
		We consider infinite lattices, which are densely connected, and show that they behave like L\'evy-flights -- or show normal behavior.
		By discussing under which conditions the effective coefficient of normal diffusion is non-finite, one can identify mechanisms of anomalous diffusion, as was done in \cite{Camboni2012}.

		The rest of the paper is structured in the following way: We first introduce Bruggeman's EMA and discuss necessary conditions on the reference model.
		In the third section we introduce the lattice L\'evy-flight and in the fourth section we discuss some examples.
		We close with discussion of the results and summary.

	\section{The effective medium approximation}
\begin{figure}
			\includegraphics[width=0.45\textwidth]{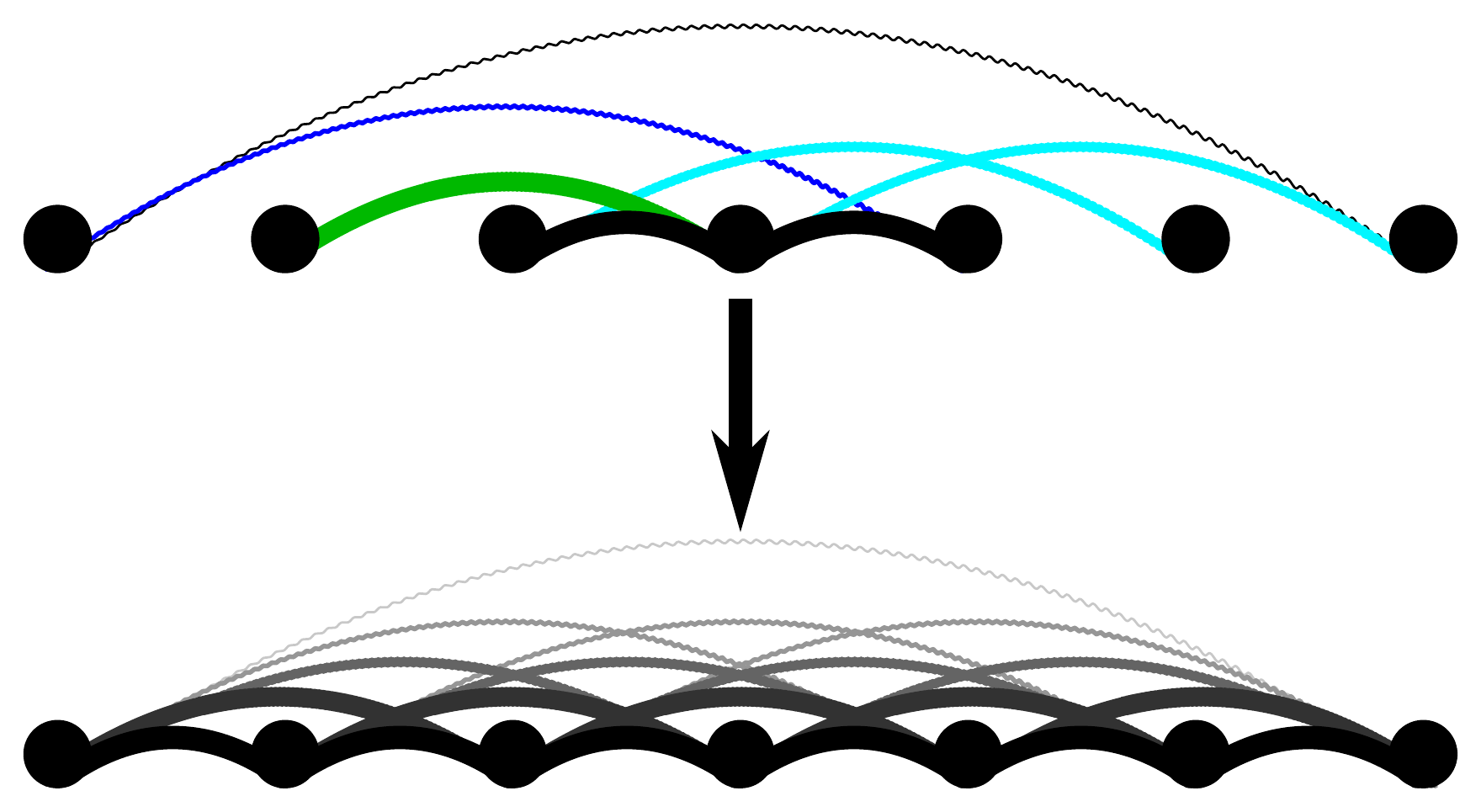}
			\caption{Sketch of EMA.
				The above depicted original lattice with long range connections is replaced by the regular one below.
				In the original lattice's bonds may be present or missing, their strength may vary randomly.
				After the EMA procedure, all bonds are present and the strength only depends on the distance of the connected lattice sites.
				\label{fig:Sketch}
			}
		\end{figure}

		Let us start with a master equation for transport in a random environment:
		\begin{equation}
				\dot{\rho}\of{x;t} 
			= 
				\br{\Laplacian\rho}\of{x;t}
			:=
				\Sum{y\in \Omega}{} w\of{x,y} 
				\brr{ \rho\of{y;t} - \rho\of{x;t} }
			.
			\label{eq:ME}
		\end{equation}
		The particles move on a lattice/graph/network $\Omega$ and jump with a symmetric rate $w\of{x,y}$ from one site to the other.
		Later we will only consider the one dimensional chain with lattice constant $a$, i.e. $\Omega = a \Integers$.
		Our arguments can be generalized to any dimension with ease, as we show in Appendix A.
		We only stick to the one-dimensional notation for simplicity and clarity.
		The environment is modeled via the transition rates $w$ that are assumed to be independently distributed for each link $\br{x,y}$.
		(In case a transition is forbidden, $w$ is put to zero.)
		We assume that the resulting graph is connected.
		We are interested in a proper averaging procedure, which replaces the random transition rates with appropriate deterministic ones.
		That means, we are looking for a deterministic function $r^*\of{x,y}$, such that
		\begin{equation}
				\dot{\rho}\of{x;t} 
			= 
				\br{\Laplacian^*\rho}\of{x;t}
			:=
				\Sum{y\in \Omega}{} \frac{ \rho\of{y;t} - \rho\of{x;t} }{r^*\of{x,y}}
			\label{eq:MERef}
		\end{equation}
		has the same qualitative (coarse-grained) behavior as the first equation.
		This will be the effective medium approximation of Bruggeman.
		We will call \eqref{eq:MERef} the ``reference model'' and Eq.\eqref{eq:ME} the ``original'' one.
		All reference quantities are denoted with stars, bare quantities belong to the disordered original model.
		$r^*\of{x,y}$ has the dimension of a time and it is in fact the mean time for the transition from $x$ to $y$.
		Later, we will identify $w\sof{x,y}$ with conductances and $r^*\sof{x,y}$ with resistances, hence the notation.
		Only for arbitrary notational reasons we are using rates in the original model and inverse rates in the reference model.
		A sketch of the replacement by the effective medium can be inspected in Fig.~\ref{fig:Sketch}.

		We think about the original model's rates as following a deterministic trend with random fluctuations, i.e. $w\of{x,y} = K\of{x,y} / f\of{x,y}$.
		Here $f\of{x,y}$ denotes the deterministic spatial dependence and $K\of{x,y}$ are independent, identically distributed random variables.
		The dimesnion of $f$ is a function of meters, e.g. $\Meter^\mu$, and $K$ has the dimension of diffusivity, i.e. $\Meter^\mu / \Sec$.
		Furthermore, we assume $f$ to be translationally invariant and isotropic, meaning it is only a function of $\Abs{x - y}$.
		Both assumptions are in no way crucial to EMA but simplify our arguments.
		As the assumption's consequence, we also split an effective generalized diffusivity $K^*$ from $r^*$ by writing: $1/r^*\of{x,y} = K^* / f^*\of{x,y}$.

		Candidates for the reference model will be lattice L\'evy flights, which have $1/r^*\of{x,y} = (a^d K^*) / \Abs{x-y}^{d+\mu}$, where $d$ is the spatial dimension.
		It will be shown that such system's behavior is only sensitive to the asymptotic behavior of $r^*$ for large $\Abs{x-y}$ (which is assumed to be a power-law).
		Hence we actually treat a whole class of disordered systems, indexed by the real parameter $\mu$.

		Our long-range jump model is very similar to the model for exciton transport on a polymer chain considered in \cite{Sokolov1997}.
		In this model, excitons can perform long jumps to far away monomers that are close in Euclidean space.
		Although, the resulting motion scales like normal diffusion, it shows paradoxical behavior as the mean squared displacement still diverges.
		The reason is a strong correlation between the shortcuts.
		Such correlations are out of the current paper's scope.
		We only investigate independent links.
		
		We repeat here two derivations of EMA that can also be found in \cite{Kirkpatrick1973} and \cite{Bustingorry2004}, as well as in textbooks \cite{Choy1999}.
		In contrast to those references, we refrain from assuming a special topology of the reference model.

		\subsection{Electrical formulation}
			The symmetry of the rates makes Eq.\eqref{eq:ME} equivalent to Kirchhoff's equations for the evolution of electric potential $\rho$ in a network of random conductances $w$ with fixed capacitances, \cite{Bouchaud1990}.
			In the stationary limit, the theory of Bruggeman proposes to replace the random conductances $w$ with deterministic effective conductances $1/r^*$ in such a way, that the average change in the stationary potential vanishes.

			Assume, $\Omega$ would be finite and we insert and extract some current on both ends of the effective medium.
			A stationary voltage profile $\rho^*$ will be assumed, as well as some stationary current.
			The effective conductance is the ratio of both.
			Afterwards, we take the thermodynamical limit and restore $\Omega$ to infinite size; the stationary current and the gradient of the potential will both vanish, but their ratio, the effective medium conductance, is correctly defined.
			This defines the effective diffusivity as the ratio of the diffusive flux and the concentration drop along a large part of the medium.
			Now we fix one link $\xi = \br{x,y}$ of the (infinite) graph, and replace the effective medium conductance $1/r^*\of{\xi}$ with the original one $w\of{\xi}$.
			This changes the voltage drop $\rho^*\of{\xi} := \rho^*\of{x} - \rho^*\of{y}$ across the link.
			To restore its former value, we introduce a current $i\of{\xi}$ on that link.
			We have 
			\begin{equation*}
					i\of{\xi}
				=
					\br{1/r^*\of{\xi} - w\of{\xi}}
					\rho^*\of{\xi}
			\end{equation*}
			The excess voltage from the replacement, $\rho - \rho^*$, can easily be computed when the total conductance $\sbr{R^*}^{-1}$ (or alternatively the total resistance $R^*$) of the lattice is known.
			The total conductance along the replaced link is $\sbr{R^*}^{-1} - \sbr{r^*}^{-1} + w$, hence we have:
			\begin{equation*}
					\rho\of{\xi}- \rho^*\of{\xi}
				=
					\frac{i\of{\xi}}{\sbr{R^*\of{\xi}}^{-1} - \sbr{r^*\of{\xi}}^{-1} + w\of{\xi}}
				.
			\end{equation*}
			We put both equations together and require that the average excess voltage vanishes.
			The average is taken with respect to the distribution of $w\of{\xi}$.
			Then the effective medium replaces the random environment faithfully.
			We obtain:
			\begin{equation}
					0
				=
					\EA{
						\frac{
							R^*\of{\xi} \br{w\of{\xi} - \sbr{r^*\of{\xi}}^{-1}}
						}{ 
							1 + R^*\of{\xi} \br{w\of{\xi} - \sbr{r^*\of{\xi}}^{-1}}
						}
					}_{w\of{\xi}}
				.
				\label{eq:EMA}
			\end{equation}
			Note that above expectation always exists, because the expression inside the brackets is bounded by unity.
			(We will show later that $R^*/r^*$ is always smaller than one.
			Hence the expression has a singularity at negative $w$, where the distribution of $w\of{\xi}$ has no support.)
			If the distribution of $w\of{\xi}$ depends on $\xi$, this procedure has to be repeated for each class of bonds and all equations have to be solved simultaneously.

			This formula has been known longer than a century and has been used many times successfully.
			Although it is not exact, it works well for systems reasonably far away from the percolation threshold.
			As we show later, the systems with long-range connections, we treat here, are always far away from the percolation threshold.
			In the short-range case, it can be augmented by the rigorous Hashin-Shtrikman bounds, \cite{Choy1999}.
			Please observe that \textit{we made no assumptions on the reference model, yet.}
			The key ingredient to solve Eq.\eqref{eq:EMA} is knowledge about the total resistance $R^*\of{x,y}$ between two nodes of the effective lattice, which is a legitimate entity from graph theory called the resistance distance, \cite{Bapat2014}.
			It is computed from the resolvent of the lattice Laplacian $\Laplacian^*$ and its definition represents an additional self-consistency requirement, since it relates $r^*$ and $R^*$.
			Thus, when the resistance distance of a certain lattice Laplacian can be found, we can use EMA to replace the random Laplacian from Eq.\eqref{eq:ME} with a deterministic one.
			To see clearer the relation between the resistance distance and the resolvent, we reformulate the approximation procedure.

		\subsection{Resolvent formulation}
			When the lattice Laplacian $\Laplacian$ in Eq.\eqref{eq:ME} has symmetric rates, we can rewrite it by sorting it by links, instead of sorting by lattice sites.
			An arbitrary order is introduced among the lattice sites and the Laplacian is written in a quasi-diagonal manner.
			This is possible if there are at most countably infinite many lattice sites.
			We use quantum mechanical notation.
			The function $\rho$ is represented by the ket $\Ket{\rho}$.
			The ket $\Ket{x_0}$ is a function that is zero everywhere but at $x$, where it is unity.
			The bra $\Bra{x_0}$ on the other hand evaluates a function at $x_0$.
			Hence, we have $\rho\of{x} = \BK{x}{\rho}$ and $\BK{x}{y} = \delta_{x,y}$.
			We write:
			\begin{align*}
					\Laplacian
				= &
					\Sum{x,y\in\Omega}{} w\of{x,y} 
					\brr{ \KB{x}{y} - \KB{x}{x} } 
				\\ = &
					\Sum{x\in\Omega}{} \Sum{y:y>x}{} w\of{x,y} 
					\brr{ \KB{x}{y} - \KB{x}{x} } 
				\\ & +
					\Sum{x\in\Omega}{} \Sum{y:y<x}{} w\of{x,y} 
					\brr{ \KB{x}{y} - \KB{x}{x} } 
				\\ = & 
					\Sum{x\in\Omega}{} \Sum{y:y>x}{} w\of{x,y} 
					\brr{ \KB{x}{y} + \KB{y}{x} - \KB{x}{x} - \KB{y}{y} }
			\end{align*}
			In the last line, we reordered the second pair of sums and interchanged the summation indices $x\leftrightarrow y$.
			Defining the new ket $\Ket{\xi} := \Ket{x} - \Ket{y}$, we have
			\begin{equation*}
					\Laplacian
				=
					\Sum{\xi\in L\of{\Omega}}{} w\of{\xi} \KB{\xi}{\xi}
				,
			\end{equation*}
			where the sum runs over all links of the lattice $\Omega$.
			One has to keep in mind, that this is not a diagonalization of the operator, since the $\Ket{\xi}$'s do not represent a basis.

			In the same way, the reference operator is reordered:
			\begin{equation*}
					\Laplacian^*
				:=
					\Sum{\xi\in L\of{\Omega}}{} 
					\frac{\KB{\xi}{\xi}}{r^*\of{\xi}}
				.
			\end{equation*}
			By taking Laplace transform of Eq.\eqref{eq:ME}, we can represent its formal solution in terms of the resolvent, or Green's operator $G\sof{s} := \sbrr{ s - \Laplacian }^{-1}$.
			Let's say we already know the resolvent $G^*\sof{s} = \sbrr{s - \Laplacian^* }^{-1}$ of the reference model.
			Then we fix a link $\xi$, and replace the effective medium conductance with the original one.
			That means we consider
			\begin{equation*}
					\Laplacian^* + \hat{D}
				:=
					\Sum{\xi'\in L\of{\Omega}}{}
					\KB{\xi'}{\xi'} 
					\brr{
						\frac{1 - \delta_{\xi,\xi'}}{r^*\of{\xi'}}
						+ 
						\delta_{\xi,\xi'} w\of{\xi} 
					}
				,
			\end{equation*}
			where $\delta_{\xi,\xi'}$ is a Kronecker-symbol.
			Finally, we express the resolvent of $\Laplacian^* + \hat{D}$ in terms of $G^*\sof{s}$.
			We use a Neumann series of $G^*\sof{s}\hat{D}$ and the fact that $\hat{D}$ has only one non-vanishing entry to write
			\begin{align*}
				&
					\BAK{x}{\brr{s - \Laplacian^* - \hat{D} }^{-1}}{y} 
					- \BAK{x}{G^*\sof{s}}{y}
				\\ = &
					\BAK{x}{\brr{ \Id - G^*\sof{s} \hat{D} }^{-1} G^*\sof{s}}{y}
					- \BAK{x}{G^*\sof{s}}{y}
				\\ = & 
					\BAK{x}{\br{
						G^*
						+
						G^* \hat{D} G^* 
						+ G^* \hat{D} G^* \hat{D} G^*
						+ \hdots
					}}{y}
					- \BAK{x}{G^*}{y}
				\\ = &
					\BAK{x}{G^*\sof{s}}{\xi} 
					\frac{\BAK{\xi}{\hat{D}}{\xi}}{1 - \BAK{\xi}{G^*\sof{s}}{\xi} \BAK{\xi}{\hat{D}}{\xi}} 
					\BAK{\xi}{G^*\sof{s}}{y}
				,
			\end{align*}
			where $\Id$ is the identity operator, and we omitted the arguments in the third line to save space.
			The EMA-requirement is that the left hand side of this equation vanishes on the average.
			We identify $\BAK{\xi}{\hat{D}}{\xi} = w\of{\xi} - \sbr{r^*\of{\xi}}^{-1}$, and take the average on both sides of the equation.
			Multiplying the equation by the factor $\BAK{\xi}{G^*\sof{s}}{\xi} / (\BAK{x}{G^*\sof{s}}{\xi}\BAK{\xi}{G^*\sof{s}}{y})$, we recover Eq.\eqref{eq:EMA}.
			We can as well identify the resistance distance in the effective medium with $\BAK{\xi}{G^*\of{s}}{\xi}$.
			In the stationary limit $t\to\infty$ (corresponding to $s\to0$ in Laplace domain), we recover the standard textbook definition of the resistance distance, see e.g. \cite{Bapat2014}:
			\begin{equation}
					R^*\of{x,y}
				:=
					\br{\Bra{x} - \Bra{y} }
					\br{\Laplacian^*}^{-1}
					\br{\Ket{x} - \Ket{y}}
				.
				\label{eq:DefRes}
			\end{equation}
			This quantity is also related to the mean first passage time from $x$ to $y$, \cite{Parris2005}.
			We see that the relation between $r^*$ and $R^*$ indeed is quite complicated, since it involves finding the Green's function.
			This may be the reason, why only certain models have been used in effective medium theory so far, in particular only \textit{short-range} models.

			Retaining the $s$-dependence in the propagator $G^*\of{s}$ leads to temporal memory in the reference system, making Eq.\eqref{eq:MERef} a generalized master equation.
			The consequences of this decision are discussed e.g. in \cite{Kenkre2009}.
			Hence above derivation is also the starting point of a non-Markovian EMA theory; this is necessary when the stationary limit is not finite.
			In this case, we do not consider $\Laplacian^*$ as a good candidate to describe the behavior of \eqref{eq:ME}.
			Here, we will rather adjust the reference model than loosing Markov-property.
			In the case of non-finite EMA diffusivities, we will say that EMA ``failed''.
			In short-range models, this is the case e.g. in the barrier model in one dimension, \cite{Bouchaud1990}.
			The existence of some extremely weak links impairs the diffusion process and leads to subdiffusion: The effective diffusivity vanishes.
			EMA also fails, when one tries to compare a long-range model with a short-range reference model.
			In this case, when the original diffusion process is superdiffusive, the reference model can not capture this feature and the effective diffusivity is infinite.
			In both cases a non-Markovian description via generalized master equations could be used, but we rather look for the correct reference model.

			In both derivations we neglected correlations between the bonds, as we only replaced \textit{one} effective link with its original.
			This renders our theory suboptimal for problems with correlated links, f.e. the site percolation problem or the random walk on a polymer chain, see ref. \cite{Sokolov1997}.
			Howeverm, EMA theories for correlated links exist as well, \cite{Yuge1977}.

			Let us now set out to investigate some of the properties of Eq.\eqref{eq:EMA}.

			\subsection{A necessary condition on the reference topology}
				Again fix one link $\xi$ and let's assume that we know the value of $R^*\of{\xi}$.
				Consider $w\of{\xi}$ and assume that with probability $c_\xi$ the link exists, i.e. $w\of{\xi} = 0$ with probability $1-c_\xi$.
				Whence, its pdf reads $\sbr{1-c_\xi}\delta\of{w} + c_\xi p_\xi\of{w}$, where $p_\xi$ denotes the pdf of the non-zero rates.
				We assume that $\Laplacian^*$ is chosen such that $1/r^*\of{\xi} = 0$ on that particular link, i.e. the link does not exist in the effective medium.
				Then, by Eq.\eqref{eq:EMA}
				\begin{equation*}
						0
					=
						0 + 
						c_\xi \Int{0}{\infty}{w} p_\xi\of{w} 
						\frac{R^*\of{\xi} w\of{\xi}}{ 1 + R^*\of{\xi} w}
					.
				\end{equation*}
				Since the integrand is positive, the integral does not vanish and this equation can not be solved, unless $c_\xi=0$.
				We conclude that, the reference model {\em must} have a link, whenever the original lattice {\em could} have a link.
				It must be chosen accordingly.
				In particular, when long-range connections are possible, i.e. when $c_\xi > 0$ for any $\xi$, short-range models have to be abandoned.
				This is the reason why classical EMA, which usually compares with a simple cubic lattice, must fail in the superdiffusive setting.
				We have to choose a long-range model as well.
				One of them, the lattice L\'evy flight is presented later.

			\subsection{Scaling transition rates and small resistance expansion}
				Let us now assume, that the rates $w\of{\xi}$ are following a deterministic spatial dependence with random fluctuations, hence $w\of{\xi} = K\of{\xi} / f\of{\xi}$ and the pdf $p_\xi$ of $w\of{\xi}$ has a scaling form.
				In this case, we can write:
				\begin{equation}
						p_\xi\of{w} 
					= 
						f\of{\xi} \tilde{p}\of{w f\of{\xi}}
					,
					\label{eq:ScalAss}
				\end{equation}
				The spatial dependence $f$ allows to define a ``bond-diffusivity'' $K\of{\xi} := w\of{\xi} f\of{\xi}$.
				The bond diffusivity does not depend anymore on the link or distance $\xi$, except for stochastic fluctuations.
				Consequently, we change the integration variable in Eq.\eqref{eq:EMA} to $K$.
				It is reasonable to also split up the reference model's transition rate by writing $1/r^*\of{\xi} = K^* / f^*\of{\xi}$.
				It is equally easy to see, that in order to solve Eq.\eqref{eq:EMA}, we have to identify $f^* = f$.
				The effective medium diffusivity $K^*$ enters as a constant factor in Eq.\eqref{eq:MERef}. 
				This factor will appear in $R^*$ as well, hence the ``reduced resistance'' $\varepsilon^*\of{\xi} := R^*\of{\xi}/r^*\of{\xi}$ is independent of $K^*$.
				Therefore, we can rewrite Eq.\eqref{eq:EMA} in terms of $K^*$ and $\varepsilon^*$:
				\begin{equation}
						0
					=
						\Int{0}{\infty}{K}
						\tilde{p}\of{K}
						\frac{
							\varepsilon^*\of{\xi}
							\br{\tfrac{K}{K^*} - 1}
						}{ 
							1 
							+ 
							\varepsilon^*\of{\xi}
							\br{\tfrac{K}{K^*} - 1}
						}
					.
					\label{eq:EMADiff}
				\end{equation}
				It is used to determine the effective diffusivity $K^*$.
				In the same line of thought we have determined the spatial dependence of the reference model's rate:
				\textit{The spatial dependence of the effective transition rates is the scaling function of the original rates's pdf.}

				A problem of the last equation is its dependence on the link $\xi$.
				This problem is resolved in two cases.
				First, when the reference topology is nearest-neighbor.
				Then there is only one class of links, $\varepsilon^*$ is a fixed number and not a function of $\xi$.
				The second solution is an asymptotic argument.
				Remember that $R^*\of{x,y}$ is the resistance between $x$ and $y$ of the total lattice, whereas $r^*\of{x,y}$ is the single resistor placed on the link $\br{x,y}$.
				Hence, $R^*$ consists of $r^*$ and possibly many other parallel resistors, consequently we can expect that $\varepsilon^* = R^* / r^* \le 1$.
				We can even expect the ratio to be much smaller than unity, when the lattice is highly connected.
				The integrand of Eq.\eqref{eq:EMADiff} is a geometric series in $\varepsilon^*\sbr{ \tfrac{K}{K^*} - 1}$, which can be expanded if $\varepsilon^*$ is sufficiently small:
				\begin{equation*}
						0
					=
						\Sum{m=1}{\infty} 
						\br{- \varepsilon^*\of{\xi} }^m
						\EA{ \br{\tfrac{K}{K^*} - 1}^m }
					.
				\end{equation*}
				As is clearly seen, another requirement of the expansion is the finiteness of all moments of $K$.
				Under these conditions, we can solve the equation for $K^*$ in leading order of $\varepsilon^*\of{\xi}$:
				\begin{equation}
						K^*
					=
						\EA{K} 
						+ \Landau{\varepsilon^*\of{\xi}}
					,	
					\label{eq:EMADiffAsym}
				\end{equation}
				which is independent on the link $\xi$.
				As we proceed to show in the next section, the reduced resistance $\varepsilon^*$ decays with distance.
				That means, we will find a positive exponent $\gamma$ such, that $\varepsilon^*\of{\xi} = \Landau{\xi^{-\gamma}}$ for large $\xi$.
				This translates \eqref{eq:EMADiffAsym} into an asymptotic statement for long ranges, because $\Landau{\varepsilon^*}$ for small $\varepsilon^*$ is then equivalent to $\Landau{\xi^{-\gamma}}$ for large $\xi$.

				The result of Eq.\eqref{eq:EMADiffAsym} is probably the most important of the whole paper and it is quite remarkable as well.
				We hereby have shown that the effective medium diffusivity for a long-range problem is equal to the arithmetic mean of the bond diffusivity.
				Note, that the arithmetic and the inverse harmonic mean diffusivity, i.e. $\sEA{K}$ and $1/\sEA{1/K}$, are the rigorous Wiener bounds for the true diffusivity of the disordered system Eq.\eqref{eq:ME}, see \cite{Choy1999}.
				Hence, we have shown that the upper bound is assumed in leading order.
				A system with long-range connections behaves like a random resistor network where all conductances are parallel.

				Eq.\eqref{eq:EMADiffAsym} also indicates that $K^*$ diverges when $\sEA{K}$ diverges as well.
				In this case, expansion of the geometric series is not allowed, and one has to solve Eq.\eqref{eq:EMADiff} directly, if possible.
				If even that is not possible, we have to declare that ``EMA failed'', that means we declare the inadequacy of the reference model.
				This inadequacy must also be declared in the inverse case, when $K^*$ vanishes, which only happens for special topologies.

				We also remark that the expansion leading to \eqref{eq:EMADiffAsym} works regardless of the scaling assumption on $w\of{\xi}$.
				Whenever $R^*$ is small compared to $r^*$, and when all the moments of the transition rates do not diverge, expansion of the geometric series in Eq.\eqref{eq:EMA} bears:
				\begin{equation*}
						\frac{1}{r^*\of{\xi}}
					=
						\EA{w\of{\xi}}
						+\Landau{R^*\of{\xi} \EA{\br{w\of{\xi} - \tfrac{1}{r^*\of{\xi}}}^2}}
					.
				\end{equation*}
				The scaling assumption is not necessary, neither is the decomposition $r^*\of{\xi} = f^*\of{\xi} / K^*$, but it can drastically reduce the complexity of the problem.

			\subsection{The coefficient of normal diffusion} 
				Let us shortly discuss, when normal diffusive behavior can be expected.
				To do so, we take the results of the last paragraphs.
				The correct reference topology has been found, the pdf of the transition rates showed scaling behavior, and we have also found the effective medium Laplacian with some $r^*\of{\xi} = f\of{\xi} / \sEA{K}$.
				We now investigate the effective coefficient of \textit{normal} diffusion, which can be defined as the time derivative of the mean squared displacement.
				When we identify the space $\Omega = a \Integers$ with a one-dimensional lattice, this quantity reads in the reference model:
				\begin{align*}
						\frac{\d}{\d t} \EA{ \br{X^*\of{t}}^2 }
					= &
						\EA{K}
						\Sum{x,\xi\in a\Integers}{} x^2
						\frac{ \rho\of{x+\xi;t} - \rho\of{x;t} }{ f^*\of{\xi} }
					\\ = &
						\EA{K}
						\Sum{x',\xi\in a\Integers}{} \rho\of{x';t} \frac{\br{x'-\xi}^2 - x'^2}{f^*\of{\xi}}
					\\ = &
						\EA{K}
						\br{
							\Sum{x'\in a \Integers}{} \rho\of{x';t}
						}
						\br{
							\Sum{\xi \in \Integers}{} \frac{\xi^2}{f^*\of{\xi}}
						}
					\\ = &
						\EA{K}
						C_2^*
					.
				\end{align*}
				We reordered the double sum in the second line.
				After that, we identified the first series with the normalization of $\rho$.
				The second series is identified as a constant that we call the ``connection factor'' $C^*_2$
				The summand proportional to $2x'\xi$ vanishes due to symmetry. 
				(We assumed isotropy: $f\of{\xi} = f\of{-\xi}$.)
				The derivation is exactly the same for higher dimensions.
				
				The connection factor tells us about the strength of the long-range connections.
				It only depends on the reference topology, and not on the fluctuations of the bond strength.
				It is zero only when all nodes of the graph are isolated.
				For a $d$-dimensional simple cubic lattice $C^*_2$ assumes the value $2d$.
				It diverges when the long-range connections are too strong, i.e. when $1/f^*\of{\xi}$ decays slower than $\xi^{-3}$.

				This equation is analogue to Eq.(2) of \cite{Camboni2012}, however without disorder in the site potentials.
				Following their rationale, we can identify mechanisms of anomalous diffusion by discussing when the coefficient of normal diffusion becomes non-finite, that means it either assumes zero or infinity.
				If this coefficient vanishes, the process is sub-diffusive; if it diverges the process must be super-diffusive.
				This is possible, when the connection-factor is zero or infinite, i.e. non-finite, or when the effective medium bond diffusivity $\sEA{K}$ is non-finite.
				We already discussed, that the connection factor may diverge for too pronounced long-range jumps.
				It vanishes only when there is no transport at all.
				The bond diffusivity $\sEA{K}$ encodes the fluctuation strength of the original transition rates.
				It may diverge, when the transition rates lack a finite first moment.
				It may vanish in a percolation-like situation, like in the one-dimensional barrier model or on any other tree structure.
				However, in the presence of long-range connections, the effective topology is nowhere close to a tree, it is rather close to a complete graph.
				Hence, the coexistence of subdiffusion due to percolation, i.e. $\sEA{K} = 0$ and superdiffusion due to long-range connections, $C^*_2 = \infty$, is impossible.
				Also, EMA is known to give horrible results in the percolation regime, as the real topology can not be compared to the effective medium topology anymore.
				In general, when the bond diffusivity is non-finite, one can use frequency dependent EMA, see e.g. \cite{Kenkre2009}, to find an effective medium description with memory.
				This description however is not Markovian anymore.
				Introducing memory results in explicit dependence on the initial state and aging, see e.g. \cite[ch. 4]{Klafter2011} or \cite{Bouchaud1992}.
				Hence, our theory reproduces the long-known truth that anomalous superdiffusion is caused by long-range jumps (diverging $C^*_2$), or by positive correlations between the steps (divergence of $\sEA{K}$, memory effects).
				Subdiffusion is only possible when $\sEA{K}$ vanishes, i.e. when the reference topology is fractal, e.g. on a percolation cluster.
				Since all considered models in this manuscript obeyed a detailed balance condition, these are mechanisms of anomalous diffusion \textit{near equilibrium}.
				In the language of \cite{Camboni2012} and \cite{Thiel2013}, this is ``structural disorder''.

				We finally turn to a possible reference model with long-range jumps.

	\section{A lattice L\'evy-flight model}
		In this model, jumps of arbitrary length are allowed, but are penalized with a power-law function.
		For a simpler notation, we only treat the one-dimensional case here.
		The computation is valid for any dimensions, though, as we show in Appendix A.
		The corresponding master equation reads:
		\begin{equation}
				\dot{\rho}\of{x;t}
			=	
				K_\mu \Laplacian_\mu \rho\of{x;t}
			:=
				K_\mu \Sum{\xi\in a\Integers}{} a
				\frac{ \rho\of{x+\xi;t} - \rho\of{x;t} }{ \Abs{\xi}^{1+\mu} }
			\label{eq:MELevy}
		\end{equation}
		Here $K_\mu$ is the anomalous diffusivity of dimension $\Meter^\mu / \Sec$ and $\mu$ is the scaling index.
		We take $\mu\in (0,2)$.
		Smaller values lead to diverging diagonal elements of the Laplacian, and would force us to only treat finite lattices; the thermodynamical limit would not be possible anymore.
		Larger values of $\mu$ suppress long jumps too much, as we have seen in the discussion of the coefficient of normal diffusion.
		The sum can be seen as a discretized version of the Riesz-Feller derivative $\sInt{\Reals}{}{\xi} \sbr{f\of{x+\xi} - f\of{x}} \sAbs{\xi}^{-1-\mu}$.
		All L\'evy-flight quantities are denoted with a $\mu$-subscript.

		The solution of the equation is given in terms of the Fourier symbol $S_\mu\of{k}$ of the operator $K_\mu\Laplacian_\mu$.
		It is defined as the operator's action in Fourier space:
		\begin{align}
				S_\mu\of{k}
			\nonumber & :=
				e^{-ikx} K_\mu \Laplacian_\mu e^{ikx}
			=
				K_\mu \Sum{\xi\in a\Integers}{} a \frac{e^{ik\xi} - 1}{\Abs{\xi}^{1+\mu}}
			\\ & =
				a K_\mu \brr{
					\mathrm{Li}_{1+\mu}\of{e^{i k a}}
					+
					\mathrm{Li}_{1+\mu}\of{e^{-i k a}}
					- 2 \zeta\of{1+\mu}
				}
			.
			\label{eq:DefSymbol}
		\end{align}
		Here $\mathrm{Li}_\alpha\of{x} := \sSum{n=1}{\infty} (x^n / n^\alpha )$ is the Polylogarithm function.
		Evaluated at $x=1$ it is equal to the Riemann-Zeta function, $\zeta\of{\alpha}$.
		The symbol can be expanded for small $k$ and the expansion gives:
		\begin{equation}
				S\of{k}
			=
				- C_\mu a^\mu K_\mu \Abs{k}^\mu
				+ \Landau{k^2}
			,
			\label{eq:SymbolAsym}
		\end{equation}
		with the positive constant $C_\mu := 2 \sAbs{ \cos\sof{\mu\pi /2} \Gamma\sof{-\mu} }$.

		Such an expansion is always possible, even when the effective medium transition rates only asymptotically behave as a power law.
		Let us assume that $1/r^*\of{\xi} = \Abs{\xi}^{-1-\mu} + \landau{\xi^{-1-\mu}}$, with some positive constant $\mu$.
		Assuming the relation is valid from some large distance, say $L$, we can split up the series at $L$, apply the asymptotic formula on one part and get:
		\begin{equation*}
				\Sum{\Abs{\xi} \le L}{} 
				\br{ e^{ik\xi} - 1 } 
				\br{ \frac{1}{r^*\of{\xi}} - \frac{1}{\Abs{\xi}^{1+\mu}} }
				+
				\Sum{\xi \in a \Integers}{}
				\frac{e^{ik\xi} - 1}{\Abs{\xi}^{1+\mu}}
		\end{equation*}
		The first sum is finite.
		A straight-forward Taylor expansion shows that it is of order $\Landau{k^2}$ for small $k$.
		The second series again gives the Polylogarithms, and restores the prior result Eq.\eqref{eq:SymbolAsym}.
		This shows, that the lattice L\'evy-flight of scaling index $\mu$ is equivalent to a whole class of processes, defined by the asymptotic behavior of $r^*$.
		Therefore, the expansion of the geometric series, performed in Eq.\eqref{eq:EMADiffAsym}, is justified in retrospect.
		In case $1/r^*\of{\xi}$ decays more rapidly than $\Abs{\xi}^{-3}$, the connection factor $\tilde{C}_2 := \sSum{\xi\in a\Integers}{} \xi^2 / r^*\of{\xi}$ converges, and the Fourier symbol grows quadratically in $k$ for small arguments.
		This, again, justifies that our main focus is $\mu \in (0,2)$.

		The resolvent of $K_\mu\Laplacian_\mu$ is given by $[ s - S_\mu\of{k}]^{-1}$ in Fourier-domain.
		From Eq.\eqref{eq:DefRes}, we see that the resistance distance is given by:
		\begin{align*}
				R_\mu\of{\xi} 
			= &
				\frac{a}{2\pi} 
				\Int{-\tfrac{\pi}{a}}{\tfrac{\pi}{a}}{k} \frac{2 - e^{ik\xi} - e^{-ik\xi}}{ - S_\mu\of{k}}
			\\ = &
				\frac{2\br{\frac{\xi}{a}}^{\mu-1}}{\pi C_\mu K_\mu}
				\Int{0}{\tfrac{\pi\xi}{a}}{\kappa} 
				\frac{ 1 - \cos\of{\kappa} }{ \kappa^\mu }
			.
		\end{align*}
		For the last equality sign, the symmetry of the integrand was exploited, Eq.\eqref{eq:SymbolAsym} was used, and finally the variable transform $\kappa := k\xi$ was applied.
		The integral converges at zero, because it is $\Landau{\kappa^{2-\mu}}$ and $\mu < 2$.
		Let us discuss its large $\xi$ behavior:
		If $1 < \mu < 2$, the integrand decays fast enough at infinity, so that the limit $\xi\to\infty$ can be taken, and the integral is finite.
		Consequently, $R_\mu\of{\xi} = \Landau{\xi^{\mu-1}}$.
		On the other hand, for $0 < \mu < 1$ the integral does not converge for $\xi\to\infty$, but grows at most as fast as $\xi^{1-\mu}$, so that $R\of{\xi}$ approaches a constant.
		For $\mu = 1$ the integral grows at most logarithmically.
		In summary, the resistance distance behaves as $R_\mu\of{\xi} = \Landau{\xi^{\Max{\mu-1}{0}}}$, but more importantly, the reduced resistance $\varepsilon_\mu := R_\mu / r_\mu$ decays:
		\begin{equation}
				\varepsilon_\mu\of{\xi}
			:=
				\frac{R_\mu\of{\xi}}{r_\mu\of{\xi}}
			=
				\Landau{\xi^{-\Min{\mu+1}{2}}}
			=
				\landau{1},
			\quad 
				\xi \to \infty
			.
			\label{eq:Eps}
		\end{equation}
		Hence, for the lattice Le\'vy-flight our asymptotic theory from the last section holds perfectly.

		All asymptotic arguments, from Eq.\eqref{eq:SymbolAsym} and forward, also hold in the normal diffusive case when $\sum \xi^2 / r^*\of{\xi}$ is finite.
		In that case, the leading order of $S^*\of{k}$ is quadratic in $k$ and, as a consequence, $R^*\of{\xi}$ grows at most linearly with distance.
		Linear growth appears, however, only in one dimension.
		
		For higher dimensions, i.e. $\Omega = a \Integers^d$, all arguments can be repeated.
		We provide these information in Appendix A.

	\section{Examples}
		In this section we discuss some example distributions $\tilde{p}\of{K}$ and compute the effective medium diffusivity $K^*$, via Eq.\eqref{eq:EMADiff}.
		We solely stick to scaling distributions.
		Three distributions where chosen: (i) the binary mixture, because Eq.\eqref{eq:EMADiff} can be solved exactly for this case.
		(ii) a power law distribution with extremely small transition rates.
		In the nearest-neighbor case, this distribution leads to subdiffusion (this is the random barrier model).
		And (iii): a Pareto-distribution.
		This distribution lacks higher moments and we show how the expansion Eq.\eqref{eq:EMADiffAsym} as well as EMA itself fail.
		\begin{figure*}
			\includegraphics[width=0.49\textwidth]{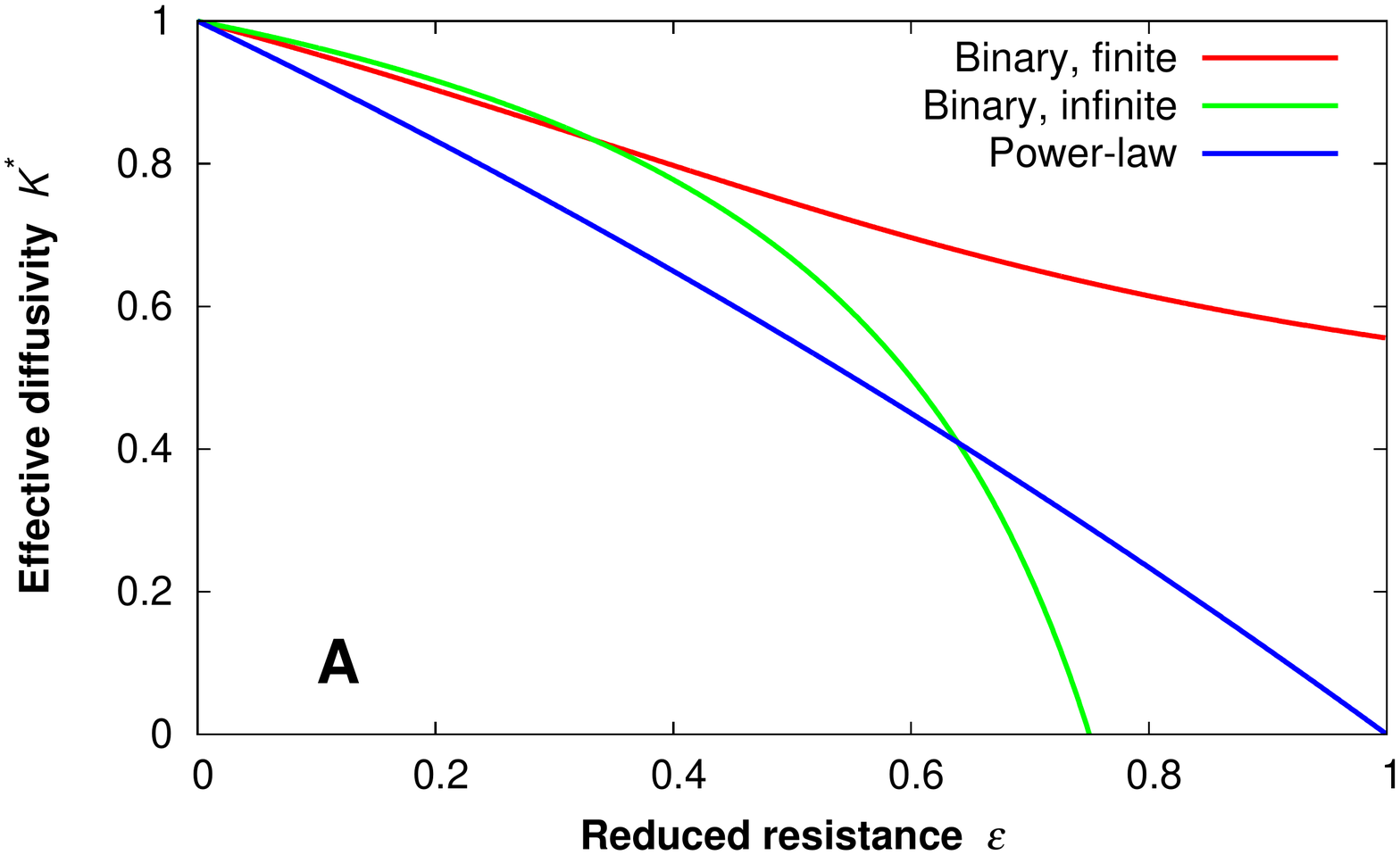}
			\hspace{-0.5ex}
			\includegraphics[width=0.49\textwidth]{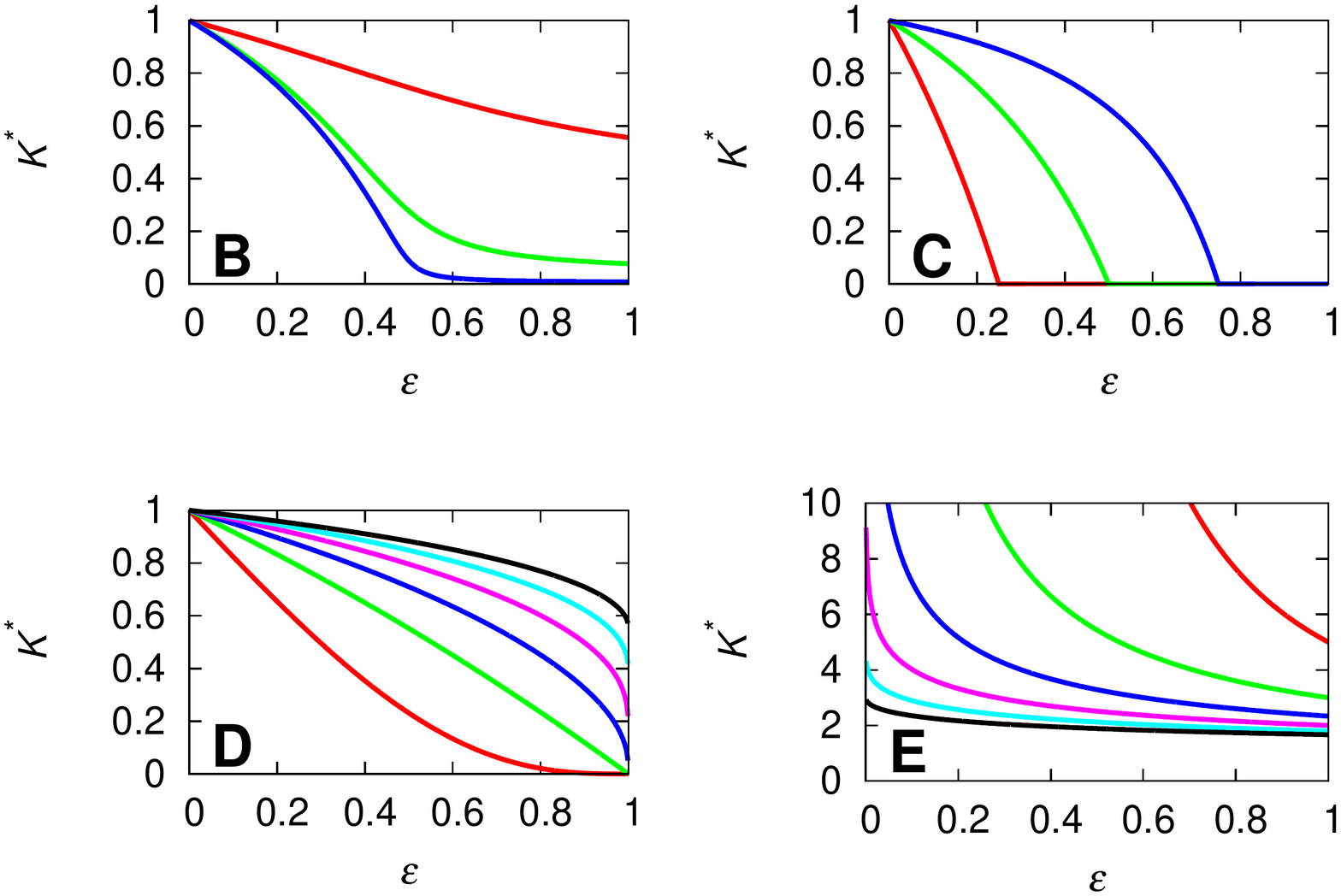}
			\caption{Effective diffusivity.
				The effective diffusivity is plotted against the reduced resistance for different distributions.
				The curves were obtained by numerical solution of equations \eqref{eq:DiffBin}, \eqref{eq:DiffIPar}, and \eqref{eq:DiffPar}.
				Except for the Pareto distribution, the limit value for $\varepsilon\to0$ is finite and served as normalization (by fixing $K_0$).
				Note the different scale for the Pareto distribution.
				(A): Comparison between different distributions: Binary distribution with $c=0.5$ and contrast $z = 5$, percolation case with $c = 0.75$ (infinite contrast), and power-law distribution with $\alpha = 0.75$.
				These distributions have been used for the simulations as well.
				(Right): All values given from top to bottom: 
				(B): Binary distribution with fixed $c=0.5$ and finite contrast $z = 5$, $50$, $500$.
				(C): Binary distribution with infinite contrast and $c=0.75$, $0.50$, $0.25$.
				The binary distribution curves converge to the curves for infinite contrast.
				(D): Power-law with $\alpha = 1.5$, $1.25$, $1.0$, $0.75$, $0.5$, $0.25$.
				(E): Pareto distribution with $\alpha = 0.25$, $0.5$, $0.75$, $1.0$, $1.25$, $1.5$.
				The effective bond diffusivity diverges here for $\alpha \le 1$, indicating the failure of our EMA-ansatz.
				\label{fig:EffDiff}
			}
		\end{figure*}

		\subsection{The binary distribution}
			For most distributions, computing the expectation in \eqref{eq:EMADiff} results in a transcendental equation for $K^*$.
			One exception is the dichotomous distribution, when the bond diffusivity is $K_0$ with probability $c$ and $K_1$ with probability $1-c$, i.e. 
			\begin{equation*}
					\tilde{p}\of{K}
				=
					c \delta\of{ K - K_0 }
					+
					\br{1-c} \delta\of{ K - K_1 }
			\end{equation*}
			Plugging this distribution into Eq.\eqref{eq:EMADiff}, leads to a quadratic equation, that can be solved for $K^*$.
			One of the solutions is negative for all $\varepsilon^*$ and can be neglected; we obtain:
			\begin{equation}
					K^*
				=
					\tilde{K}\of{\varepsilon^*} 
					+ \sqrt{ \tilde{K}\of{\varepsilon^*}  + \frac{\varepsilon^*}{1-\varepsilon^*} K_0 K_1 }
				,
				\label{eq:DiffBin}
			\end{equation}
			with 
			\begin{equation*}
					\tilde{K}\of{\varepsilon^*}
				=
					\frac{1}{2} \brr{ 
						\frac{c - \varepsilon^*}{1-\varepsilon^*} K_0
						- \frac{1 - c - \varepsilon^*}{1 - \varepsilon^*} K_1
					}
				.
			\end{equation*}
			Some remarks are in place:
			First, in the limit $\varepsilon^* \to0$ we recover Eq.\eqref{eq:EMADiffAsym} and have $K^* = \EA{K} = cK_0 + \sbr{1-c}K_1$.
			Corrections are of order $\Landau{\varepsilon^*\of{\xi}}$, and vanish for large $\xi$.
			Secondly, we can consider the percolation problem by setting $K_1$ to zero.
			This gives
			\begin{equation}
					K^*
				=	
					\frac{c - \varepsilon^*}{1-\varepsilon^*} K_0
				,
			\end{equation}
			and shows that the percolation threshold for this problem is $c_\text{cr}=\varepsilon^*$, consequently $c_\text{cr} = 0$ in the asymptotic limit.
			This means the system always percolates, which is not surprising, considering the highly connected topology in the lattice L\'evy-flight.
			This result may be contrasted with the result for $d$-dimensional simple cubic systems.
			Here, $\varepsilon^* = R^*/r^* = 1/d$ and EMA predicts $c_\text{cr} = 1/d$.

			In Fig.~\ref{fig:EffDiff} the effective diffusivity is plotted against the reduced resistance $\varepsilon^*$.
			The upper left panel shows the result of the dichotomous distribution for different contrasts $z := K_1 / K_0$.
			As the contrast increases, one class of bonds becomes negligible and the curves tend to the ones of the ``percolation'' case, depicted in the upper right panel.
			However, in both cases $K^*$ assumes a finite value as $\varepsilon^*$ approaches zero.
			Keep in mind, that only the value of $K^*$ for $\varepsilon^*\to0$ matters, since all corrections can be asymptotically neglected.
			They will not alter the diffusive behavior.

		\subsection{Power-law distribution}
			The next distribution is a power-law one:
			\begin{equation*}
					\tilde{p}\of{K}
				=
					\frac{\alpha}{K_0} \br{ \frac{K}{K_0} }^{\alpha - 1} 
					\Theta\of{K - K_0} \Theta\of{-K}
				,
			\end{equation*}
			where $\Theta\of{x}$ is the Heavyside step function.
			If $K$ is distributed like this, $1/K$ has a Pareto-distribution. 
			To compute the expectation in Eq.\eqref{eq:EMADiff}, we first use the identity $\tfrac{x}{1+x} = 1 - \tfrac{1}{1+x}$, then we identify the remaining integral with a hypergeometric one, \cite{Abramowitz1972}.
			The effective diffusivity is defined by the implicit equation
			\begin{equation}
					0
				=
					1 - \frac{1}{1 - \varepsilon^*} \HGF{- \frac{\varepsilon^*}{1 - \varepsilon^*}\frac{K_0}{K^*}}{\alpha}{1}{\alpha+1}
				.
				\label{eq:DiffIPar}
			\end{equation}
			In the one-dimensional next-neighbor model with $\alpha < 1$, this distribution leads to subdiffusion as it lacks a harmonic mean $\EA{1/K}$, and the mean transition time diverges.
			Like higher dimensional models, the long-range model retains its diffusive properties.
			The roots of this equation neither diverge nor vanish, the reason is topology.
			Broken or very weak links can easily be avoided.

			The results from the last equation are shown in the lower left panel of Fig.~\ref{fig:EffDiff} for different values of $\alpha$.
			As in the dichotomous distribution, $K^*$ approaches a finite value as $\varepsilon^*$ approaches zero.

		\subsection{Pareto distribution}
			The last example is a Pareto-distribution for the transition rates:
			\begin{equation*}
					\tilde{p}\of{K}
				=
					\frac{\alpha }{K_0} \br{ \frac{K}{K_0} }^{-\alpha - 1} 
					\Theta\of{K - K_0}
				.
			\end{equation*}
			We consider this case, because this distribution lacks moments of higher order than $\alpha$.
			Consequently, the expansion that leads to Eq.\eqref{eq:EMADiffAsym} is prohibited for $\alpha < 1$.
			We can not approximate the effective diffusivity with the average transition rate, as the latter diverges.
			Using the same steps as before, together with the variable transformation $z := 1/K$, we find a similar hypergeometric function that defines $K^*$:
			\begin{equation}
					0
				=
					1 - \frac{\alpha}{\alpha + 1} \frac{K^*}{\varepsilon^* K_0}
					\HGF{- \frac{1 - \varepsilon^*}{\varepsilon^*}\frac{K^*}{K_0}}{\alpha+1}{1}{\alpha+2}
				.
				\label{eq:DiffPar}
			\end{equation}
			For $\alpha \le 1$, numerical inversion of this equation shows a divergence at $\varepsilon^* \to 0$, indicating the failure of our EMA model.
			In this case, the lattice L\'evy-flight is not the correct reference model to describe the diffusion process.
			Two possible ways are open to deal with this problem: finding a better reference Laplacian, i.e. modifying Eq.\eqref{eq:MELevy}, or introducing memory, and loosing the Markov property in the description.

			The numerical inversion for the Pareto distribution is shown in the lower right panel of Fig.~\ref{fig:EffDiff}.
			Since the distribution has no mean value for $\alpha < 1$, $K^*$ diverges as $\varepsilon^*\to0$, and effective medium theory can not be used to replace the original system with a lattice L\'evy flight.
			Here the diffusive behavior is indeed altered by the distribution of the transition rates.
			In fact, the fluctuations of the individual elements overwhelm the deterministic spatial dependence of the transition rates.

	\section{Numerical verification}
		Effective medium theory's main purpose is quantitative applicability.
		Therefore, we shortly discuss how to perform numerical experiments of Eq.\eqref{eq:ME} and how to determine the effective quantities.

		\subsection{Simulation scheme}
			For a fixed random environment, i.e. the set of all transition rates $w\of{x,y}$, Eq.\eqref{eq:ME} describes a random walk.
			The random walker's probability, $p\of{x,y}$, to jump from $x$ to another lattice site $y$ and its the mean sojourn time, $\tau\of{x,y}$ are given by:  
			\begin{equation*}
					p\of{x,y}
				=
					\frac{w\of{x,y}}{\Sum{z\in\Omega}{} w\of{x,z}}
				, \quad
					\tau\of{x}
				=
					\frac{1}{
						\Sum{z\in\Omega}{} w\of{x,z}
					}
				.
			\end{equation*}
			The time needed for this transition is an exponentially distributed random variable whose mean is the inverse of the sum of rates.
			This way one obtains the random walker's trajectory $X\of{t}$ in the random environment.
			The procedure is repeated for many samples of the environment, and averages are taken from such an ensemble, whence the average is taken with respect to environment \textit{and} thermal history of the walker.

		\subsection{Measuring the effective quantities}
			When the reference model is known, validity of the effective medium can be tested by investigating the characteristic function of the random walk.
			It is the expectation value $\sEA{\exp\sof{ikX\of{t}}}$, i.e. Fourier-transform of the pdf of $X\of{t}$.
			The pdf of $X\of{t}$ is also the propagator, consequently the characteristic function is the inverse Laplace transform of $G^*\sof{s} = \sbrr{s - S^*\of{k}}^{-1}$, which is an exponential.
			In case the random walk is symmetric, the odd component of the expectation vanishes and it suffices to take the cosine function $\sEA{\cos\sof{kX\of{t}}}$.
			Finally taking the logarithm of the characteristic function results in a linear law in time:
			\begin{equation}
					\ln\EA{ e^{ikX\of{t}} }
				=
					\ln\EA{ \cos\of{kX\of{t}} }
				=
					S^*\of{k} t
				.
				\label{eq:CharFuncFit}
			\end{equation}
			Fitting the logarithm of the empirical characteristic function against time reveals $S^*\of{k}$.

			When the functional form of the symbol is known, parameters like $K^*$ or $\mu$ can be fitted from the result.
			For the L\'evy-flight this would be an asymptotic power-law fit.
			The problem is to determine a good range of wave-vectors for the fit.
			
			Using the characteristic function is necessary in the case of long-range connections, because the mean squared displacement is no longer finite.
			Lower moments of fractional order are hard to access analytically.
			Therefore we followed the idea of \cite{Postnikov2015}, to inspect the characteristic function.

			\begin{figure*}
				\includegraphics[width=0.49\textwidth]{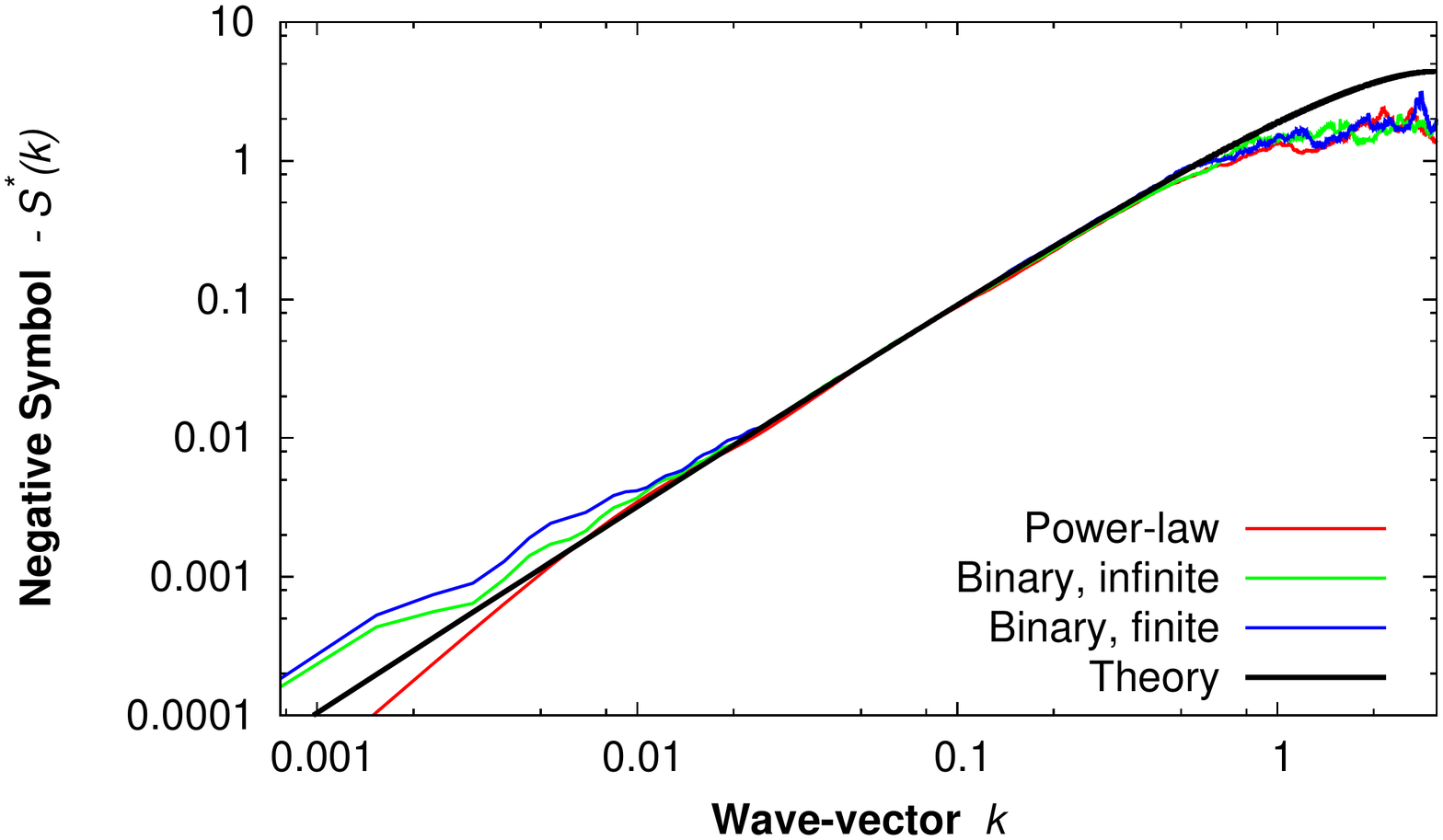}
				\includegraphics[width=0.49\textwidth]{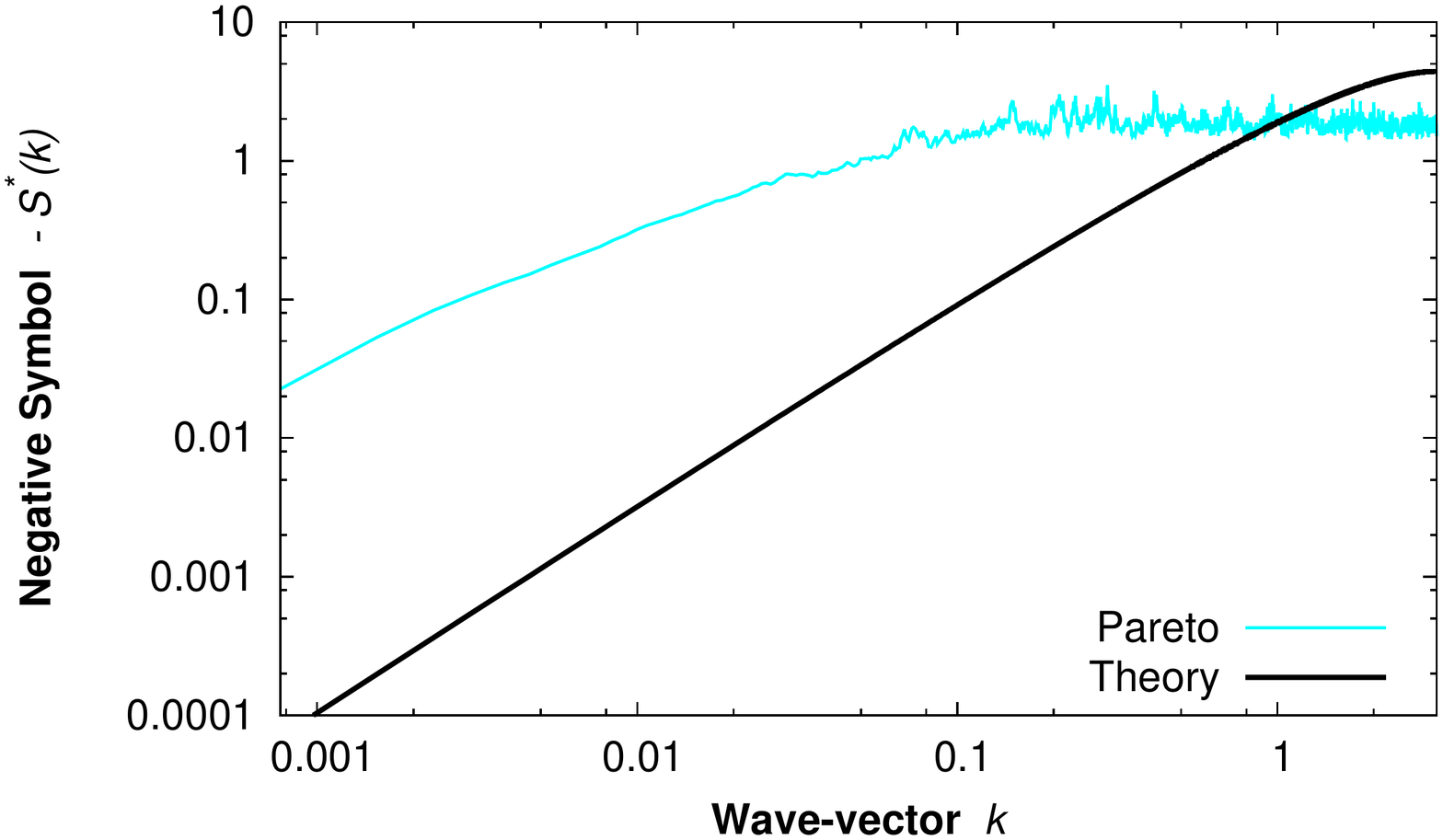}

				\caption{Measurements of the Fourier-symbol.
					The symbol was determined from numerical simulations like described in the main text.
					We performed simulations with $\mu=1.5$ for the binary distribution with $c = 0.5$ and contrast $z = 5$, for percolation with $c=0.75$, as well as for the power-law distribution with $\alpha=0.75$ (all left).
					The random walk with Pareto-distributed transition rates ($\alpha = 0.75$) is shown on the right hand side.
					The theory curve is the prediction from Eq.\eqref{eq:DefSymbol}.
					The left hand plot shows great agreement for intermediary values of $k$, spanning two orders of magnitude from $k=0.01$ to $k=1$.
					Better agreement for small $k$ can be achieved by using larger lattices, better agreement for large $k$ is achieved by considering smaller total simulation times.
					EMA fails for the Pareto distribution (right hand side), hence this curve does not match the theory, neither in offset nor in slope.
					\label{fig:Symbol}
				}
			\end{figure*}

		\subsection{Pitfalls}
			Several practical problems appear, when one performs the numerics.
			All of them are due to the finiteness of the lattice.
			First of all, one should keep in mind that the highest possible resolution of wave-vectors is $k = 2\pi m / L$, where $L$ is the length of the lattice and $m$ is an integer.
			Secondly, the random walker quickly enters stationary state.
			Due to long-range jumps, equilibration essentially starts with the first or second jump.
			In practice, the interesting small $k$-behavior gets pushed to the origin and can not be resolved properly, when the observation time is too large.
			
			Finally, the Fourier symbol of the finite lattice differs from the infinite lattice's symbol.
			In a finite lattice, the symbol is a finite sum; ultimately it behaves like $k^2$ for small wave-vectors.
			Therefore, an asymptotic expansion of the empiric symbol may not be captured the asymptotic expression given by Eq.\eqref{eq:SymbolAsym} for very small $k$.
			The best range for fitting the symbol hence is an intermediary, not too small and not too large.
			All effects become worse for smaller $\mu$.

		\subsection{Results of the numerical investigation}
			We performed Monte-Carlo simulations as prescribed above using ensembles of $2048$ random walkers (each in its own environment sample), on a lattice with $L = 8192$ sites and lattice constant $a = 1$.
			Each trajectory was recorded $129$ times until final time $4$.
			The symbol was inferred from a linear fit of equation Eq.\eqref{eq:CharFuncFit}, evaluated for wave vectors $k = 2\pi m /L$, with an integer $0<m<L/2$.
			We simulated all above discussed examples; the result can be seen in Fig.\ref{fig:Symbol}.

			We found that for large $k$ the symbol enters a noisy regime, which can be shifted to the right when smaller final times are considered.
			This is to be expected, since the characteristic function decays exponentially in time, making it harder to obtain larger values of the symbol.

			It can be seen that except for the Pareto distribution, all curves fall on the prediction, given by Eq.\eqref{eq:DefSymbol}.
			This can be seen best in the double-logarithmic plot on the right hand side.
			However, due to finite size effects, this agreement only holds for intermediary values of $k$.
			We refrained from fitting the symbol, since the diffusivities have already been normalized to unity.
			As expected, the Pareto curve is way off, and has a lower slope than the other curves, indicating a much faster motion than in the other examples.

	\section{Summary and discussion}
		EMA puts us in the position of replacing a randomly disordered diffusion system, with a deterministic reference model.
		We worked out two restrictions on the reference model: 
		Firstly, it has to possess a link, wherever the original model could have a link.
		And secondly, the spatial decay of the reference model's transition rates is given by the scaling of the original rates' pdf, provided the mean transition rates are finite.
		With those rules, EMA can be applied with any reference network, for which the propagator (or the resistance distance in the static case) is known.
		By inspecting the effective coefficient of normal diffusion, we also discussed some mechanisms of anomalous diffusion.
		The resistance distance for the lattice L\'evy-flight was computed and several example distributions were discussed.
		We showed that the predictions obtained from EMA excellently agree with random walk simulations.

		EMA's main advantage from the analytical point of view is the ability to treat a disordered system equivalent to a translationally invariant one.
		The emergence of translational invariance is a consequence of homogenization: 
		At large enough length scales, a disordered system behaves like an ordered, homogeneous one.
		This enables the analytician to use Fourier transform (which was our main tool to derive the theory of lattice L\'evy-flights).
		Hopefully this technique will be employed in the mean field description of more complex problems with random long-range connections, like synchronization or infection spreading, \cite{Tessone2006,Cencini2008,DaSilva2013,Kuo2015}.

		In this paper, we focused on infinite systems, when the term ``free diffusion'' makes sense.
		Then a proper thermodynamic limit can be taken.
		This rules out the small world and scale-free networks considered in \cite{Parris2005,Candia2007,Parris2008}, for they do not penalize long jumps with a distance factor, as we did.
		EMA can be applied for any finite graph, but taking the lattice size to infinity can result in diverging diagonal elements of the effective medium Laplacian.
		We showed that EMA can also fail and bear a vanishing or diverging effective bond diffusivity $K^*$.
		Although a zero effective bond diffusivity is impossible in the long-range case, divergence is indeed possible, as was shown with the Pareto distribution.
		The correct choice of the reference model is still an open question.
		The authors' educated guess is that the Pareto distribution may not admit a reference model with proper thermodynamical limit.

	\section*{Acknowledgments}
		This work was funded by DFG within IRTG 1740 research and training group project and within project SO 307/4-1.

	\appendix
	\section{Higher dimensional lattice L\'evy-flights}
		In this appendix, we derive asymptotic expressions for the Fourier symbol and the resistance distance in arbitrary dimensions.
		The Laplacian and the master equation take the form:
		\begin{equation*}
				\dot{\rho}\of{\V{x};t}
			=
				K_{\mu,d} a^\mu \Sum{\V{y} \in a \Integers^d}{} 
				\frac{ \rho\of{\V{y};t} - \rho\of{\V{x};t} }{\Abs{\V{x}-\V{y}}^{d+\mu}}
			.
		\end{equation*}
		The Fourier symbol of this operator is given by 
		\begin{equation*}
				S_{\mu,d}\of{\V{k}}
			=
				K_{\mu,d} a^\mu
				\Sum{\V{\xi} \in a \Integers^d}{} 
				\frac{e^{i\V{k}\V{\xi}} - 1}{\Abs{\V{\xi}}^{d+\mu}}
		\end{equation*}
		Unfortunately, this series can not easily expressed in terms of some special function, as was the case for one dimension.
		Instead, we will approximate the series with an integral and switch to spherical coordinates.
		The approximation can be justified by e.g. the Euler-MacLaurin formula.
		\begin{align*}
				S_{\mu,d}\of{\V{k}}
			= &
				K_{\mu,d} 
				\Int{\Sphere^{d-1}}{}{\V{n}} \Int{0}{\infty}{\xi} \xi^{d-1} 
				\frac{e^{i\xi\V{k}\V{n}} - 1}{\xi^{d+\mu}}
			\\ = &
				K_{\mu,d} 
				\Abs{\V{k}}^\mu
				\Int{\Sphere^{d-1}}{}{\V{n}} \Int{0}{\infty}{\kappa} 
				\frac{e^{i \kappa \V{e}_{\V{k}}\V{n}} - 1}{\kappa^{1+\mu}}
			\\ = & 
				- C_{\mu,d} K_{\mu,d} \Abs{\V{k}}^\mu
			.
		\end{align*}
		Here, the outer integral is the angular integration over the $(d-1)$-sphere.
		In the second line we changed the integration variable to $\kappa := \xi \Abs{\V{k}}$.
		The integrand decays faster than $\kappa^{-1}$ for large $\kappa$.
		By isotropy the integrand behaves like $\kappa^{1-\mu}$ for small $\kappa$ (the first order vanishes when the angular integration is performed).
		Since $\mu < 2$, the double integral converges and we call it $- C_{\mu,d}$.
		Higher order corrections are introduced by rigorous application of Euler-MacLaurin formula.

		Let us now turn to the resistance distance.
		It is given by the integral over the Brillouin zone $\mathbb{B} := [-\pi/a, \pi/a)^d$:
		\begin{equation*}
				R_{\mu,d}
			=
				\br{\frac{a}{2\pi}}^d \Int{\mathbb{B}}{}{\V{k}} 
				\frac{2 - 2\cos\of{\V{k}\V{\xi}}}{ - S_{\mu,d}\of{\V{k}}}
			.
		\end{equation*}
		As before this integral converges at $\V{k} = 0$.
		We will proceed to show that it grows slower than $\Abs{\V{\xi}}^{d+\mu}$, hence, that the reduced resistance decays, just like in the one dimensional case.
		Note first, that the integrand is non-negative for all $\V{k}$, hence we can enlarge the integration domain from a cube to a sphere with radius $\sqrt{d}\pi / a$ to derive an upper bound for $R_{\mu,d}$.
		Then again, we switch to spherical coordinates $\V{k} =: \V{n} k$ and introduce $\kappa := k \Abs{\V{\xi}}$.
		We obtain:
		\begin{equation*}
				R_{\mu,d}
			<
				\frac{2 \Abs{\V{\xi}}^{\mu-d}}{C_{\mu,d} K_{\mu,d}}
				\br{\frac{a}{2\pi}}^d 
				\Int{\Sphere^{d-1}}{}{\V{n}} \Int{0}{\frac{\sqrt{d}\pi \Abs{\V{\xi}}}{a}}{\kappa}
				\frac{1 - \cos\of{\kappa \V{n}_{\V{\xi}} \V{n}}}{ \kappa^{\mu + 1 - d} }
			.
		\end{equation*}
		The large $\Abs{\V{\xi}}$-behavior of the remaining integral is limited by the power factor $\kappa^{d - 1 - \mu}$.
		The integral converges when $\mu > d$.
		Otherwise it grows at most like $\Abs{\V{\xi}}^{d - \mu}$.
		Together with the prefactor we have
		\begin{equation*}
				R_{\mu,d}
			< 
				\Landau{\Abs{\V{\xi}}^{\Max{0}{\mu - d}}}
			,
		\end{equation*}
		and for the reduced resistance we obtain:
		\begin{equation*}
				\varepsilon_{\mu,d}\of{\V{\xi}}
			=
				\Landau{\Abs{\V{\xi}}^{- \Min{\mu+d}{2d}}}
			.
		\end{equation*}

	\bibliographystyle{aipnum4-1}
	\bibliography{Article,Book,Self,NotRead}
\end{document}